\documentclass[twocolumn,showpacs,pra]{revtex4-1}
\usepackage{graphicx}
\usepackage{amsfonts,amssymb,amsmath}
\usepackage{color}

\def\br{{\bf r}}
\def\KS{\text{KS}}
\def\ext{\text{ext}}
\def\inte{\text{int}}
\def\barr{\text{barr}}
\def\kin{\text{kin}}
\def\step{\text{step}}
\def\C{\text{C}}
\def\H{\text{H}}
\def\X{\text{X}}
\def\HXC{\text{HXC}}

\begin{document}
\title{Kohn-Sham potential for a strongly correlated finite system with fractional occupancy}
\author{A. Ben\'itez}
\altaffiliation[Present address:\;]{Catalan Institute of Nanoscience and Nanotechnology (ICN2), 
Campus UAB, Barcelona 08193, Spain}
\affiliation{Centro At\'{o}mico Bariloche and Instituto Balseiro, Comisi\'{o}n Nacional
de Energ\'{\i}a At\'{o}mica, 8400 Bariloche, Argentina}
\author{C. R. Proetto}
\email[Electronic address:\;]{proetto@cab.cnea.gov.ar}
\affiliation{Centro At\'{o}mico Bariloche and Instituto Balseiro, Comisi\'{o}n Nacional
de Energ\'{\i}a At\'{o}mica, 8400 Bariloche, Argentina}

\date{\today}

\begin{abstract}
Using a simplified one-dimensional model of a diatomic molecule, the associated interacting density and 
corresponding Kohn-Sham potential have been obtained analytically for all fractional molecule occupancies $N$
between 0 and 2. For the homonuclear case, and in the dissociation limit, the exact Kohn-Sham potential
builds a barrier at the midpoint between the two atoms, whose strength increases linearly with $N$, with
$1 < N \leq 2$. In the heteronuclear case, the disociating KS potential besides the barrier also exhibits a plateau 
around the atom with
the higher ionization potential, whose size (but not its strength) depends on $N$.
An anomalous zero-order scaling of the Kohn-Sham potential with regards to the strength of the electron-electron
repulsion is clearly displayed by our model; without this property both the unusual barrier and plateau features
will be absent.
\end{abstract}

\maketitle

\section{Introduction}
Ground-state density-functional theory (DFT) maps the interacting electronic problem into an effective
non-interacting system, that shares with the real system the ground-state density and energy~\cite{PY89,DG90}.
In the Kohn-Sham (KS) formulation of DFT~\cite{KS65}, the electrons are acted on by an effective single-particle, multiplicative
potential, the KS potential. All the complicated many-body effects of the real system are fully included in
this effective potential, whose crucial role is to do whatever is needed to reproduce the density and total 
energy of the 
real interacting system. The aim of the present work is to explore some of the unusual features of the 
exact KS potential, in particular its behavior when the system under study has a non-integer (fractional) number of
electrons.

There are several reasons why the analysis of systems with fractional charges within the DFT framework may be of
some interest. A good example is the behavior of the molecule 
$H_2^+$, when stretched by increasing the distance between the two protons beyond the equilibrium separation.
At the dissociation limit, and without any symmetry breaking, $H_2^+$ correctly splits in two identical fragments or open subsystems,
each one consisting of a proton and half of one electron: $H^{+0.5} \cdots H^{+0.5}$. 
Most of the local or semilocal energy functionals of common use in DFT behave poorly in this limit of fractional occupancies, yielding energies
far below the proper binding energy of 1 Hartree, due to the tendency of approximate functionals to spread out the 
electron density artificially. In practical calculations, the problem may be ``solved'' by 
breaking the spatial symmetry and localizing 
the electron on one of the two protons. Proceeding this way, the resulting binding energy of the $H_2^+$ molecule
at its dissociation limit is reasonable,
but the associated density is not correct. 
In exact DFT, on the other side, the problem is solved by the condition that the energy of the two
hydrogen-like fragments must be the same either if the electron is localized in one of the protons, or if 
there is half of one electron at each proton~\cite{PRCVS07}.
The same happens with all radical symmetric
molecules $A_2^+$ at infinite bond length. This failure has been denoted the ``many-electron self-interaction
error''~\cite{RPCVSS07,SCY06*} or ``delocalization error''~\cite{SCY08} of semilocal functionals, 
and happens when some occupied KS orbitals share an
electron between two open subsystems, equivalent to having in each subsystem a non-integer occupation number.
A related situation occurs with many asymmetric molecules $AB$ that with approximate functionals dissociate not to neutral atoms $A$ and $B$
but improperly to fractionally charged fragments $A^{+q} \cdots B^{-q}$~\cite{RPCVSS07}. 
Also, long-range charge transfers are usually
overestimated~\cite{T03}, and the energy barriers that control the reaction rates in chemical reactions are underestimated
or even absent~\cite{PSTS08}. All these problematic issues of current DFT may be related to specific missing features of the 
KS potential resulting from semilocal functionals.

From a rigorous point of view, the extension of the ground-state DFT formalism to the case of systems whose density integrates
to a fractional number was made in a seminal work by Perdew {\it et al.}~\cite{PPLB82}. They proved, 
by introducing a zero-temperature ensemble
DFT formalism, that the total energy is a piecewise linear function of the electron number between two adjacent
integers. In turn, this leads to the theoretical prediction of a discontinuity of the KS potential as the electron
number passes through an integer. This abrupt jump of the exact KS potential, missed in all local or semilocal
approximations, is on the other side the crucial ingredient that explains the severe underestimation of the 
fundamental band-gap of insulators and semiconductors~\cite{PL83,SS83}. All these issues have been discussed in
the influential review of Ref.~[\onlinecite{P85}]; for a more updated review, see Ref.~[\onlinecite{CSW12}].

The aim of this work is to illustrate some of the non-intuitive features of the KS potential through the use of a
simple but strongly correlated one-dimensional model of the hydrogen molecule, for which the exact interacting
density is available for any electron number between 0 and 2. By a reverse-engineering procedure, the $N$-dependent 
exact KS potential is then obtained from the $N$-dependent density. The simplicity of our model allow us to prove
unambiguously an anomalous scaling of the correlation potential, with regard to the strength of the electron-electron
repulsion. The present work is organized as follows: in Section II we introduce the model and explain the 
method we use for finding its solutions; in Section III we provide the main numerical and analytical results,
while Section IV is devoted to the Conclusions.


\section{Model and method of solution}
The bottleneck of our reverse-engineering method is in finding the solution (in principle exact) of a $N=2$ closed shell  diatomic molecule.
As this is not easily available, and we are more interested in understanding the physics behind the KS potential than 
in describing a real tridimensional molecule, we will simplify the model. Following Ref.~[\onlinecite{HTR09}],
we will use the following one-dimensional mimic of a diatomic molecule (in atomic units (a.u.))
\begin{eqnarray}
 \left[-\frac{1}{2} \left(\frac{\partial^2}{\partial x_1^2}+\frac{\partial^2}{\partial x_2^2} \right)
 +v_{\ext}(x_1)+v_{\ext}(x_2) \right. \nonumber \\
 \big. +v_{\inte}(|x_1-x_2|) \bigg] \Psi^{\gamma}(x_1,x_2) = E(2) \Psi^{\gamma}(x_1,x_2) \; ,
 \label{2eSe}
\end{eqnarray}
with
\begin{equation}
 v_{\ext}(x) = -v \; [\lambda \; \delta(x+d/2)+\delta(x-d/2)] \; ,
 \label{vext}
\end{equation}
and
\begin{equation}
 v_{\inte}(|x-x'|) = \gamma \; \delta(x-x') \; .
 \label{contact}
\end{equation}
As we will see, this simplified model preserves all the main physical ingredients of the real three-dimensional
molecule with the long-range Coulomb interactions.
Here, $v$ and $\lambda$ are both positive; $\lambda=1$ ($\neq 1$) corresponds to the homonuclear (heteronuclear) diatomic 
molecule. Since a one-dimensional attractive delta-potential has only one bound state~\cite{G96},
the two attractive delta potentials in $v_{\ext}(x)$ provides the two-dimensional bound-states basis needed for the 
forthcoming considerations.
$v_{\ext}(x)$ represents in our model the attractive Coulomb potentials from the two protons of the real hydrogen
molecule.
Eq.~(\ref{contact}) represents the repulsive interaction ($\gamma \geq 0$) between the two electrons in the molecule, expressed
here by a short-range (contact) interaction. The non-interacting limit of Eq.~(\ref{2eSe}) is obtained for $\gamma=0$,
while $\gamma \rightarrow \infty$ drives the ``molecule'' towards the strongly-interacting limit. 
The value of the ratio $v/\gamma$ moves the system from the weakly interacting regime ($v/\gamma \gg 1$) towards
the strongly interacting regime ($v/\gamma \ll 1$).
And the nice point is that the ground-state $\Psi^{\gamma}(x_1,x_2)$ and $E(2)$
may be found analytically for $\gamma \rightarrow \infty$, by appealing to the boson-fermion mapping~\cite{YG05}. 
More precisely,
\begin{equation}
 \Psi^{\gamma \rightarrow \infty}(x_1,x_2) = |\phi_+(x_1) \phi_-(x_2) - \phi_+(x_2) \phi_-(x_1)| \; ,
 \label{bfm}
 \end{equation}
and $E(2) = \varepsilon_+ + \varepsilon_-$.
In equation above, the symbol $|...|$ represents the absolute value, and $\phi_{\pm}(x)$ and $\varepsilon_{\pm}$ are
the normalized eigenfunctions and eigenvalues of the following single-particle Schr\"odinger equation,
\begin{equation}
 \left[-\frac{1}{2}\frac{\partial^2}{\partial x^2} + v_{\ext}(x)\right] \phi_{\pm}(x) = \varepsilon_{\pm} \, \phi_{\pm}(x) \; .
 \label{1eSe}
\end{equation}
From Eqs.~(\ref{1eSe}) and (\ref{vext}) one obtains
\begin{equation}
 \phi_{\pm}(x) = A_{\pm}\left( e^{-\alpha_{\pm}|x+d/2|} + f_{\pm} e^{-\alpha_{\pm}|x-d/2|} \right) \; ,
 \label{wf}
\end{equation}
with
\begin{equation}
 A_{\pm}^2 = \frac{\alpha_{\pm}}{1+f_{\pm}^2+2f_{\pm}e^{-d\alpha_{\pm}}(1+d\alpha_{\pm})} \; ,
\end{equation}
\begin{equation}
 \alpha_{\pm} = \frac{v(1+\lambda)}{2} \pm \frac{v}{2} \left[ (1-\lambda)^2+4\lambda e^{-2d\alpha_{\pm}} \right]^{1/2} ,
 \label{alphapm}
\end{equation}
\begin{equation}
 f_{\pm} = \frac{ve^{-d\alpha_{\pm}}}{\alpha_{\pm}-v} \; ,
\end{equation}
and $\varepsilon_{\pm}=- \; \alpha_{\pm}^2/2$; since $\alpha_+ \geq \alpha_-$, $\varepsilon_+ \leq \varepsilon_-$.
The density of the two-electron system is defined by
\begin{equation}
 \rho_2(x) := \int \left[\Psi^{\gamma \rightarrow \infty}(x,x')\right]^2 dx' = \phi_+^2(x) + \phi_-^2(x) \; ,
 \label{rho2}
\end{equation}
which should be contrasted with the non-interacting two-electron density $\rho_2^0(x)=2 \, \phi_+^2(x)$.

Following the prescription of zero-temperature ensemble DFT~\cite{PY89,DG90}, the ground-state density for the diatomic molecule with
$N$ electrons, with $0 \leq N \leq 2$, is given by
\begin{eqnarray}
 \rho_N(x) = \left\{
 \begin{array}{l l}
  N\rho_1(x) & \quad \text{if $0 \leq N \leq 1$}, \nonumber \\
  \\
  (2-N) \rho_1(x) + (N-1) \rho_2(x) & \quad \text{if $1 \leq N \leq 2$}, \\
  \end{array} \right. \\
  \label{density*}
\end{eqnarray}
with $\rho_2(x)$ defined in Eq.~(\ref{rho2}), and $\rho_1(x)=\phi_+^2(x)$~\cite{convexity}. Replacing we obtain that
\begin{eqnarray}
 \rho_N(x) = \left\{
 \begin{array}{l l}
  N \phi_+^2(x) & \quad \text{if $0 \leq N \leq 1$}, \nonumber \\
  \\
  \phi_+^2(x) + (N-1) \phi_-^2(x) & \quad \text{if $1 \leq N \leq 2$} \; . \\
  \end{array} \right. \\
  \label{density**}
\end{eqnarray}
This result for the interacting density should be contrasted with the $N$-dependent non-interacting density
$\rho_N^0(x) = N \phi_+^2(x)$, for $0 \leq N \leq 2$. Both $\rho_N(x)$ and $\rho_N^0(x)$ are identical for 
$0 \leq N \leq 1$, as it should be. However, for $N>1$, while $\rho_N^0(x)$ continues with the progressive 
occupancy of the single-particle ground-state orbital $\phi_+(x)$, $\rho_N(x)$ places the fractional occupancy beyond unity fully in the 
first-excited orbital $\phi_-(x)$. This striking difference between both densities is displayed in Fig.~1 for the 
homonuclear case ($\lambda=1$), and in Fig.~2 for the heteronuclear case ($\lambda \neq 1$).

\begin{figure}[h]
\begin{center}
\includegraphics[width=8.6cm]{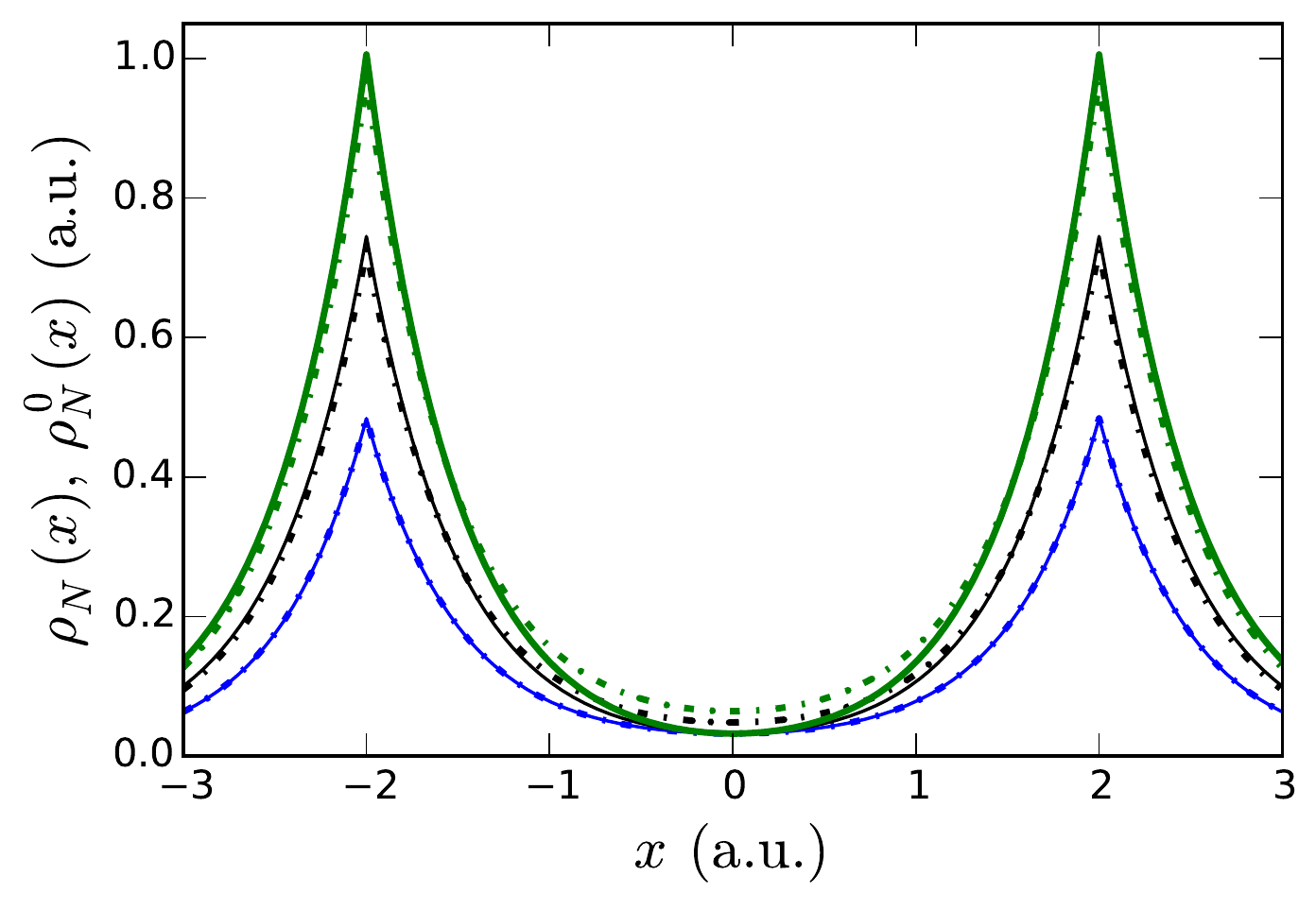}
\caption{$\rho_N^0(x)$ (dotted lines) and $\rho_N(x)$ (full lines) for the homonuclear case, and 
$N=1~\text{(blue)},\;1.5~\text{(black)},\;2~\text{(green)}$. $v=\lambda=1$, and $d=4$.
For $N=1$ both densities coincide.}
\label{Fig1}
\end{center}
\end{figure}

\begin{figure}[h]
\begin{center}
\includegraphics[width=8.6cm]{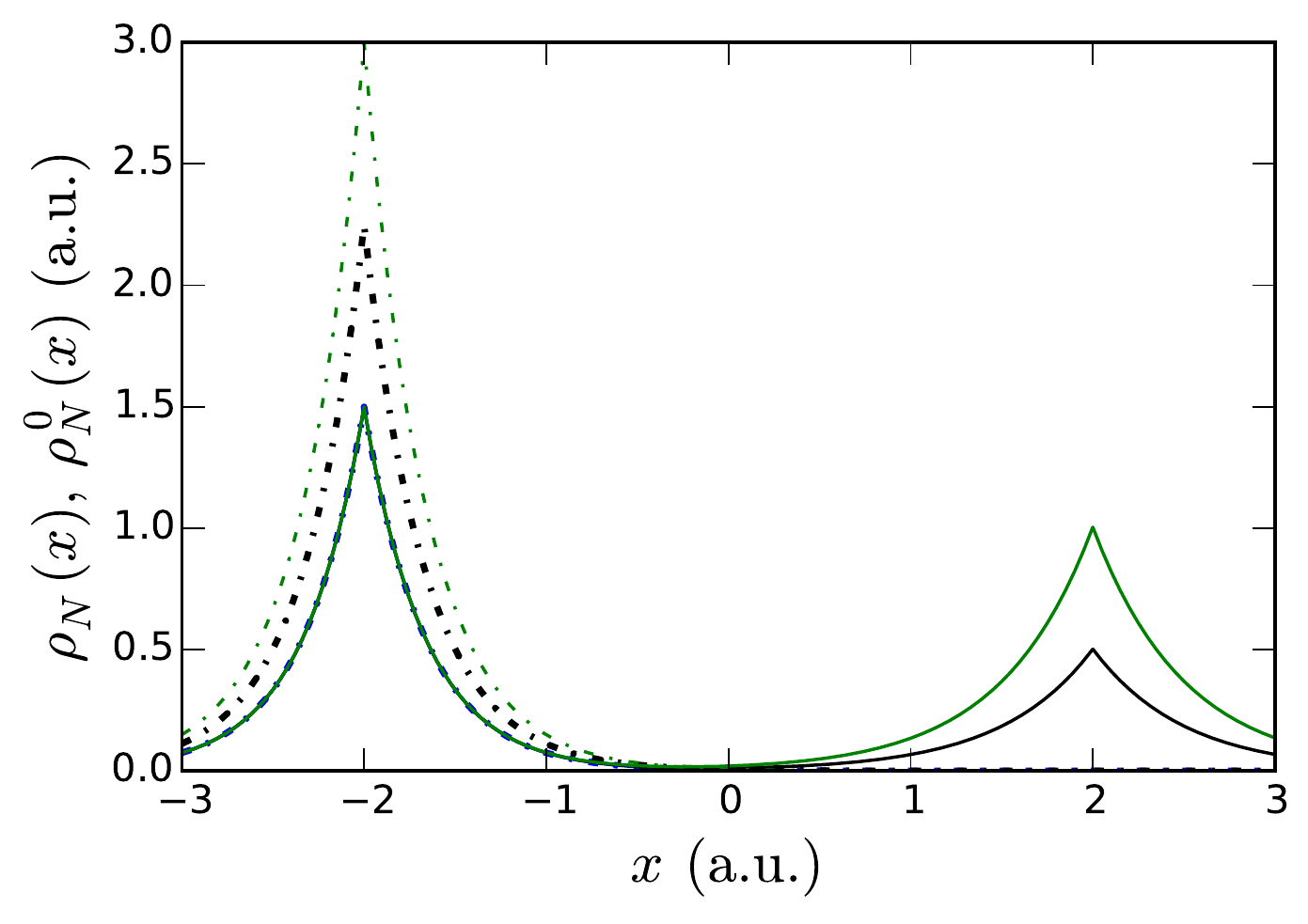}
\caption{$\rho_N^0(x)$ (dotted lines) and $\rho_N(x)$ (full lines) for the heteronuclear case, and 
$N=1~\text{(blue)},\;1.5~\text{(black)},\;2~\text{(green)}$. $v=1, \; \lambda=1.5$, and $d=4$.
For $N=1$ both densities coincide.}
\label{Fig2}
\end{center}
\end{figure}

Having $\rho_N(x)$, the KS potential is obtained by reverse engineering from the expression
\begin{equation}
 v_{\KS}^N(x) = \frac{1}{2\sqrt{\rho_N(x)}}\frac{\partial^2\sqrt{\rho_N(x)}}{\partial x^2} + C \; ,
 \label{KSp*}
\end{equation}
with $C$ being a constant to be fixed later. While Eq.~(\ref{KSp*}) is self-evident in the KS framework for
$N=1$ and $N=2$, its applicability in the full range $0 \leq N \leq 2$ has been discussed and validated in
Ref.~[\onlinecite{SP08}].
Replacing the $N$-dependent density $\rho_N(x)$ in Eq.~(\ref{KSp*}), it yields
\begin{eqnarray}
 v_{\KS}^N(x) = \left\{
 \begin{array}{l l}
  v_{\ext}(x) + \Delta v_{\KS}^<(x) &  \text{if $0 \leq N \leq 1$}, \nonumber \\
  \\
  v_{\ext}(x) + \Delta v_{\KS}^>(x) &  \text{if $1 < N \leq 2$}, \\
 \end{array} \right. \\
\end{eqnarray}
where $\Delta v_{\KS}^<(x) = \varepsilon_+ + C^<$,
$\Delta v_{\KS}^>(x) = v_{\C}^{\barr}(x) + v_{\C}^{\step}(x) + C^>$, and
\begin{equation}
 v_{\C}^{\barr}(x) = \frac{(N-1)\left[\phi_+^{'}(x)\phi_-(x) - \phi_-^{'}(x)\phi_+(x)\right]^2}{2[\rho_N^>(x)]^2} \; ,
 \label{Cp*}
\end{equation}
\begin{equation}
 v_{\C}^{\step}(x) = - \frac{\varepsilon_+\phi_+^2(x)+(N-1)\varepsilon_-\phi_-^2(x)}{\rho_N^>(x)} \; ,
 \label{Cp}         
\end{equation}
with $\rho_N^>(x)$ equal to $\rho_N(x)$ for $N \geq 1$, and primes denoting derivate with respect to the coordinate $x$.
For $N=2$, $\Delta v_{\KS}^>(x)$ reduces to the expression obtained in Ref.~[\onlinecite{HTR09}] for the same model.
Choosing $C^<=- \, \varepsilon_+$, we obtain that $\Delta_{\KS}^<(x) \equiv 0$. 
In the asymptotic limit $|x| \gg d$, $\phi_+(x)/\phi_-(x) \rightarrow 0$, 
$v_{\C}^{\barr}(x)$ goes to zero, while $v_{\C}^{\step}(x)$ approaches $- \, \varepsilon_-$; as a consequence 
$\Delta_{\KS}^>(x \gg d) \rightarrow - \, \varepsilon_- + C^>$.
Choosing $C^>=\varepsilon_-$, $\Delta_{\KS}^>(|x| \gg d) \rightarrow 0$.
$\Delta v_{\KS}^>(x)$ has the following interesting property: $\lim_{|x| \rightarrow \infty} \Delta v_{\KS}^>(x) \rightarrow 0$,
but $\lim_{|x| \rightarrow \infty} \lim_{N \rightarrow 1^+} \Delta v_{\KS}^>(x) \rightarrow \varepsilon_- - \varepsilon_+ = 
I(1)-A(1) > 0$~\cite{PPLB82}. Here we have defined $I(N)=E(N-1)-E(N)$ and $A(N)=I(N+1)=E(N)-E(N+1)$ as the 
ionization potential and electronic affinity, respectively, of the $N$-electron molecule. This is precisely the 
discontinuity of the $N$-dependent KS potential addressed above, when crossing integer values of $N$ ($N=1$ in this case).
Using a different model with an external harmonic confinement, Ref.~[\onlinecite{GGS09}] analyzes the discontinuity
for the case $N=1$, while in Ref.~[\onlinecite{GT14}] the discontinuity has been analyzed for several real atoms,
for other values of $N$.

The case $N=2$ of present model for the diatomic molecule has been generalized in Ref.~[\onlinecite{HTR09}], 
replacing the delta-function
potentials in Eqs.~(\ref{vext}) and (\ref{contact}) by soft-Coulomb potentials of the type $1/\cosh^2(x)$. The 
results between the two models are quite similar, particularly in the dissociation limit where
the respective KS potentials become essentially identical.

\section{Results and discussions}

\subsection{Homonuclear case}
For $\lambda=1$, the eigenvalue equation simplifies to $\alpha_{\pm}=v(1 \pm e^{-d\alpha_{\pm}})$, and this leads to
$f_{\pm}= \pm 1$, and $A_{\pm}^2 = (\alpha_{\pm}/2)/[1 \pm e^{-d\alpha_{\pm}}(1+d\alpha_{\pm})]$.
$\phi_+(x)$ becomes the symmetric ``bonding'' solution, while $\phi_-(x)$ becomes the antisymmetric 
``antibonding'' solution. 
In the dissociation limit, defined as $vd \gg 1$, $\alpha_+ \sim \alpha_- \sim v$, and since 
$\varepsilon_- \sim \varepsilon_+ \sim - \, v^2/2$, the ``step'' contribution to $\Delta v_{\KS}^>(x)$ becomes a constant that
cancels with $C^> (= \varepsilon_-)$. For the ``barrier'' contribution one obtains that if $-d/2 < x < d/2$,
\begin{equation}
 v_{\C}^{\barr}(x,vd \gg 1) \simeq 
 \frac{4I(N-1)}{[2-N+N\cosh(2\sqrt{2I}x)]^2} \; ,
 \label{vbarr}
\end{equation}
with $I=v^2/2$ being the ionization potential of any of the two identical fragments.
Clearly $v_{\C}^{\barr}(x)$ vanishes for $N=1$ and has the largest value for $N=2$, while as a function of
the coordinate $x$ it has the shape of a barrier centered at $x=0$, of height $I(N-1)$. 
We display in Fig.~3 a drawing of $ v_{\C}^{\barr}(x)$ for two typical situations,
using the expression in Eq.~(\ref{Cp*}).

The physics behind this barrier in the KS potential is best understood by writing the interacting density for
$1 \leq N \leq 2$ as
\begin{equation}
 \rho_N^>(x) = \rho_N^0(x) + (N-1)\left[\phi_-^2(x)-\phi_+^2(x)\right] \; .
 \label{density}
\end{equation}
As may be appreciated from Fig.~1, the difference $\phi_-^2(x)-\phi_+^2(x)$ is positive for 
$x \sim \pm \; d/2$, but negative in the ``bonding'' region $x \sim 0$. Considering both facts,
one concludes that the main difference between $\rho_N^0(x)$ and $\rho_N(x)$ is that the electronic
charge (even when fractional) tends to be more localized around the ``atoms'' in the latter, as a way of
diminishing the repulsive interaction. And the only way that the KS effective one-body potential has to provide a
ground-state density equal to $\rho_N(x)$ is through the building of a barrier around the ``bonding'' region, as
given by Eq.~(\ref{vbarr}).
It is interesting to note that once the dissociation limit $vd \gg 1$ sets in, $ v_{\C}^{\barr}(x,vd \gg 1)$ is
independent on the distance $d$ between the two atoms. The size of the barrier, on the other side, depends
crucially on an atomic property of the two separated fragments, the ionization potential $I$,
and also on the number of electrons in the molecule.

\begin{figure}[h]
\begin{center}
\includegraphics[width=8.6cm]{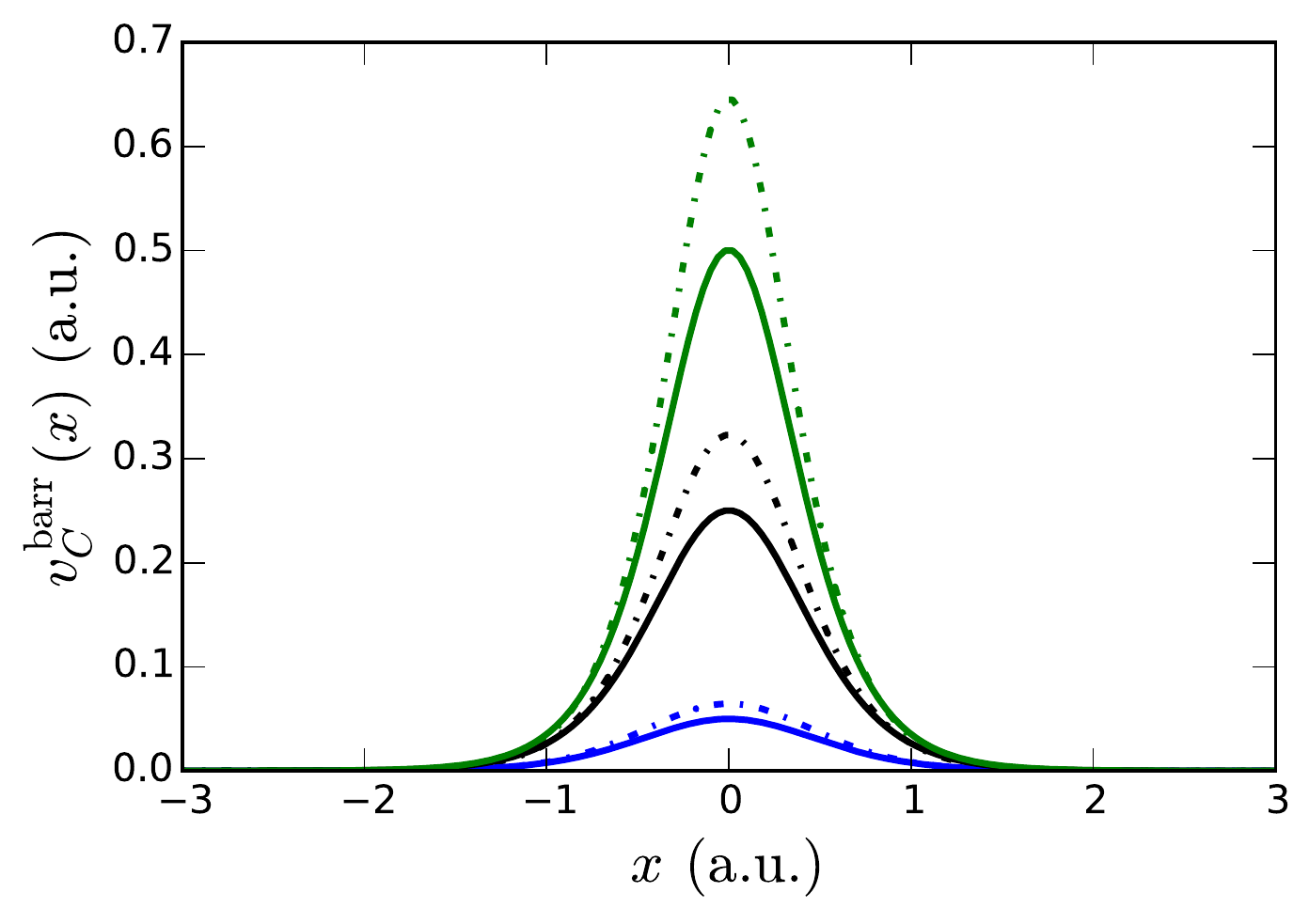}
\caption{$v_{\C}^{\barr}(x)$ for $N=1.1~\text{(blue)}, 1.5~\text{(black)}, 2~\text{(green)}$. $v=\lambda=1$, and $d=4$ (dotted lines) and 10 (full lines).}
\label{Fig3}
\end{center}
\end{figure}

From Eq.~(\ref{vbarr}) one obtains that 
$v_{\C}^{\barr}(0,vd \gg 1) = I(N-1) \simeq 0.05$, 0.25, and 0.50 for $N$= 1.1, 1.5, and 2, respectively, and for 
the parameters in Fig.~3. This confirms that for $d=10$, $v_{\C}^{\barr}(x)$ is fully in the dissociation limit,
but not when $d=4$. When $vd \gg 1$, the electronic density in the bonding region is
exponentially small ($\sim e^{-dv}$), and then the only role of the barrier consists in maintain the difference
$\phi_-^2(x) - \phi_+^2(x)$ positive for $x \sim \pm \; d/2$. Its second role, depressing the density in the interatomic
region becomes irrelevant in the dissociation limit, since as discussed above the density is exponentially small there.
As a consequence, in the dissociation limit the barrier becomes essentially an atomic property, independent of the 
distance between the fragments, and with an ``intrinsic'' height of $I(N-1)$. 
When the two atoms become closer ($d=4$ in Fig.~3), the height of the barrier increases beyond its dissociation
limit, since it becomes more difficult to isolate the electronic charge of the two approaching fragments.

Returning to Eq.~(\ref{density**}), it is seen that in the dissociation limit it can be written as
\begin{eqnarray}
 \rho_N^>(x,vd \gg 1) &\sim& \frac{N}{2}[\rho_L(x)+\rho_R(x)] \nonumber \\ 
 &+& (2-N) \sqrt{\rho_L(x)} \sqrt{\rho_R(x)} \; ,
 \label{entanglement}
\end{eqnarray}
with $\rho_L(x)=\sqrt{2I_L} e^{-2\sqrt{2I_L}|x+d/2|}$, $\rho_R(x)=\sqrt{2I_R} e^{-2\sqrt{2I_R}|x-d/2|}$
being the normalized densities associated to the left and right fragments, respectively. In the approximation
of Eq.~(\ref{entanglement}), $I_L \simeq I_R \simeq I$.
It shows that for $1 \leq N < 2$, the interacting density is not the plain sum of the density of the left
and right fragments. Only for $N=2$, $\rho_2(x,vd \gg 1) \sim \rho_L(x)+\rho_R(x)$, as found in Ref.~[\onlinecite{HTR09}].
Note that if, erroneously, one would approximate $\rho_N^>(x,vd \gg1)$ by the sum of the two atomic densities,
the resulting $v_{\KS}^N(x)$ would be independent of $N$~\cite{note*}, contrary to the rigorous result for $v_{\C}^{\barr}(x,vd \gg 1)$
in Eq.~(\ref{vbarr}). In other words, when electronic systems with fractional charges are involved, the
hallmark of the dissociating limit is not always the fact that the density can be written as the sum of the
isolated or atomic densities.

For $N=2$, the appearance of a potential barrier at the midpoint between the two protons in the $H_2$
molecule has been discussed long ago, in works by Baerends and coworkers~\cite{GB97,BBS89}. 
What we have denoted
here as $v_{\C}^{\barr}(x)$ is what these authors denoted as $v_{\kin}(\br)$. Since the definition of this
``kinetic'' potential is in terms of a conditional probability, which in turn is defined only for integer occupancies
(it will be $N=2$ in our case), the latter may be considered as a particular case of the former.

\subsection{Heteronuclear case}
As can be appreciated from Fig.~2, the evolution of $\rho_N^>(x)$ is quite different from the 
$\rho_N^>(x)$ for the homonuclear case displayed in Fig.~1, while also exists marked differences between the 
heteronuclear densities $\rho_N^>(x)$ and $\rho_N^0(x)$. The differences are easily seen in the dissociation
limit $e^{-vd} \ll |1-\lambda|$; expanding the square root in Eq.~(\ref{alphapm}), it yields
\begin{equation}
 \alpha_{\pm} \simeq \frac{v(1+\lambda)}{2} \pm \frac{v|1-\lambda|}{2} 
                \left[ 1 + \frac{2 \lambda e^{-2d\alpha_{\pm}}}{(1-\lambda)^2} \right] \; .
\end{equation}
To proceed, let us choose that $\lambda > 1$. One obtains then that 
$\alpha_+(\lambda > 1) \sim v\lambda $, $\alpha_-(\lambda>1) \sim v - v \lambda e^{-2vd} / (\lambda-1) $,
$f_+(\lambda>1) \sim e^{- v \lambda d} / (\lambda -1) \ll 1$, and 
$f_-(\lambda>1) \sim (\lambda-1) e^{vd}/ \lambda \gg 1$.
Returning to Eq.~(\ref{wf}), if $f_+(\lambda > 1) \ll 1 $, this implies that in the dissociating limit
$\phi_+(x)$ essentially corresponds to $\phi_L(x)$ ($=\sqrt{\rho_L(x)}$), while $\phi_-(x)$ is essentially equal to 
$\phi_R(x)$ ($=\sqrt{\rho_R(x)}$).
This already explains the results in Fig.~2: a) since $\rho_N^0(x) = N \phi_+^2(x) \sim N \phi_L^2(x)$, increasing
$N$ the density of the left-well increases, while the right-well remains essentially empty; b) since
$\rho_N^>(x) = \phi_+^2(x)+(N-1)\phi_-^2(x) \simeq \phi_L^2(x)+(N-1)\phi_R^2(x)$, as $N$ increases beyond unity
now the density at the right-well is the one that increases, while the density at the left-well remains 
essentially constant, and equal to its value at $N=1$; c) since $\alpha_+ > \alpha_-$, the decay of the density
around $x=-\,d/2$ is faster than the decay of the density around $x=d/2$.
Choosing instead $\lambda < 1$, the roles of $\phi_L(x)$ and $\phi_R(x)$ are exchanged, resulting in an interacting
density, for instance, given by $\rho_N^>(x) = \phi_R^2(x) + (N-1)\phi_L^2(x)$.

This is the way that the interacting system finds to minimize the repulsive contact interaction: for $0 \le N \le 1$ we 
have a non-interacting system and all the charge goes to the single-particle orbital with the lowest energy; for $1 < N \le 2$ the
interacting system locates {\it all} the extra charge in the opposite well.

\begin{figure}[h]
\begin{center}
\includegraphics[width=8.6cm]{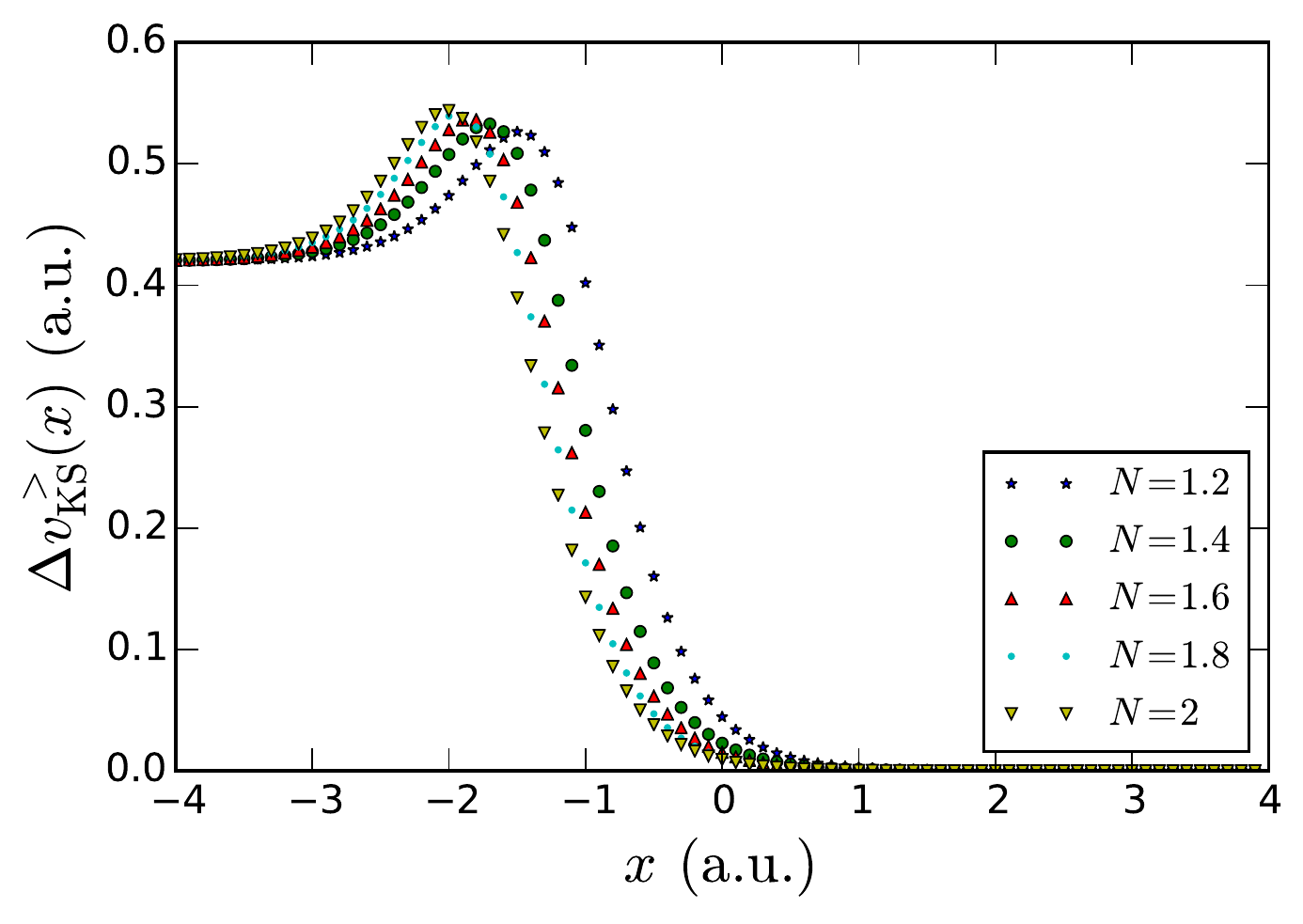}
\caption{$\Delta v_{\KS}^{>}(x)$ for $N=1.2, 1.4, 1.6, 1.8, 2$. $v=0.4$, $\lambda=2.5$, and $d=10$.}
\label{Fig4}
\end{center}
\end{figure}

We display in Fig.~4 $\Delta v_{\KS}^>(x)$, for $d$ large enough such that the molecule
is in the dissociation limit. The mapping $\phi_+(x) \rightarrow \phi_L(x)$  ($\phi_R(x)$) 
and $\phi_-(x) \rightarrow \phi_R(x)$  ($\phi_L(x)$) for $\lambda > 1 \, (\lambda <1)$ in the dissociation limit
provides us with alternative expressions for $v_{\C}^{\barr}(x, vd \gg 1 )$ and $v_{\C}^{\step}(x, vd \gg 1)$, as follows
\begin{equation}
 v_{\C}^{\barr}(x, vd \gg 1) = \frac{(N-1)(\sqrt{2I_L}+\sqrt{2I_R})^2\rho_L(x)\rho_R(x)}{2[\rho_L(x)+(N-1)\rho_R(x)]^2}
                            , \label{maxover}
\end{equation}
and
\begin{equation}
 v_{\C}^{\step}(x, vd \gg 1) =  \frac{I_L\rho_L(x)+(N-1)I_R\rho_R(x)}{\rho_L(x)+(N-1)\rho_R(x)} \; .
\end{equation}
Replacing the expressions for $\rho_L(x)$ and $\rho_R(x)$ in the equations above,
one obtains after some manipulation
\begin{equation}
 v_{\C}^{\barr}(x,vd \gg 1) = \frac{(\sqrt{2I_L}+\sqrt{2I_R})^2/8}{\cosh^2\left[(\sqrt{2I_L}+\sqrt{2I_R})(x+x_0)\right]} \;,
 \label{barr}
\end{equation}
while
\begin{equation}
 v_{\C}^{\step}(x,vd \gg 1) = \frac{I_L-I_R}{1+\exp{\{2[(\sqrt{2I_L}+\sqrt{2I_R})(x+x_0)]}\}} \;,
 \label{step}
\end{equation}
where
\begin{eqnarray}
 x_0(N) &=& \frac{1}{2(\sqrt{I_L}+\sqrt{I_R})} \nonumber \\
        &\times&  \left[ d(\sqrt{I_L}-\sqrt{I_R}) + \frac{1}{\sqrt{2}}
        \ln\left(\frac{\sqrt{I_R}(N-1)}{\sqrt{I_L}} \right) \right] \; . \nonumber \\
 \; .
 \label{x0}
\end{eqnarray}
Note that the dependence on $N$ is hidden now in the parameter $x_0(N)$, that signals the location of the 
peak feature.
For $N=2$, these equations reduce to the ones found in Ref.~[\onlinecite{HTR09}]. These equations are only valid
for $-d/2 \leq x \leq d/2$, and $I_L > I_R$. For $I_L = I_R$ and $N=2$, $x_0(2) = 0$ and 
$v_{\C}^{\step}(x,vd \gg 1) \equiv 0$; only in this case Eq.~(\ref{barr}) coincides with the homonuclear
result of Eq.~(\ref{vbarr}) for $v_{\C}^{\barr}(x,vd \gg 1)$. For any other $N$ between 1 and 2, 
$v_{\C}^{\step}(x,vd \gg 1)$
still vanishes for $I_L=I_R$, but the correct homonuclear limit for $v_{\C}^{\barr}(x,vd \gg 1)$ as given by
Eq.~(\ref{vbarr}) cannot be obtained as a limit from Eq.~(\ref{barr}). This is quite reasonable, as the starting
point for the derivation of Eqs.~(\ref{barr},\ref{step}) above is that in the dissociation limit the density can be
written as $\rho_N^>(x)=\rho_L(x) + (N-1)\rho_R(x)$, with $\rho_L(x),\rho_R(x)$ being the densities of the left and right 
wells,
or viceversa. And such simplification for the interacting density is not possible (except for $N=2$), as can be
seen from Eq.~(\ref{entanglement}). The explanation for this apparently paradoxical situation is given below.

The structure ``step'' + ``barrier/shoulder'' observed in Fig.~4 has been already discussed previously~\cite{P85,HTR09},
but its dependence on $N$, to the best of our knowledge has been not studied before. Using the 
parameters corresponding to Fig.~4, one obtains that $x_0(1.2) \simeq 1.24$, $x_0(1.4) \simeq 1.49$,
$x_0(1.6) \simeq 1.63$, $x_0(1.8) \simeq 1.74$, $x_0(2) \simeq 1.82$, in agreement with the position of the 
peak in the shoulder, which is located at $x = - x_0(N)$. Also, $I_L-I_R = (\alpha_L^2-\alpha_R^2)/2 \simeq 0.42$.

The physics behind the barrier/shoulder and step features in $\Delta v_{\KS}^>(x)$ have been already discussed for $N=2$, 
and here we
provide additional insight by considering the case of fractional $N$. The crucial concept is that 
the effective single-particle potential $\Delta v_{\KS}^>(x)$ should manage to reproduce the interacting ground-state
density, as displayed in Fig.~2. The step structure forms around the ``atom'' with the higher ionization potential
(the left atom in Fig.~2), and locally induces the ``equilibrium'' condition 
$\varepsilon_L + I_L - I_R = \varepsilon_R$, which allows the exclusive population of the right-well in Fig.~2,
as soon as $N>1$. The barrier/shoulder structure comes from $v_{\C}^{\barr}(x)$, and as in the homonuclear case its role is
provide a barrier that induces the decoupling of the two atoms, in the dissociation limit. 
This can be seen clearly from Eq.(\ref{maxover}): $v_{\C}^{\barr}(x,vd \gg 1)$ will display its maximum strength
at the coordinate $x$ where the left and right densities have its maximum overlap. in ``units'' of the density
$\rho_N(x)$. By inspection of Eq.~(\ref{maxover}) is easy to check that this happens when
$\rho_L(x) = (N-1)\rho_R(x)$. Solving this equation for $x$, one obtains Eq.~(\ref{x0}) in an alternative way.
As $N$ increases beyond 1, the effective right-related density $(N-1)\rho_R(x)$, and the point of maximum overlap
between the two atomic density distributions moves closer to the atom with the higher ionization energy, as 
seen in Fig.~4. These considerations also explain why the shoulder/barrier peak moves closer to the left well as
$N$ increases in Fig.~4.

Proceeding analogously, the following expressions are obtained for $x \leq -d/2$,
\begin{equation}
 v_{\C}^{\barr}(x,vd \gg 1) = \frac{(\sqrt{2I_L}-\sqrt{2I_R})^2/8}{\cosh^2\left[(\sqrt{2I_L}-\sqrt{2I_R})(x+x_0')\right]} \;,
 \label{barr*}
\end{equation}

\begin{equation}
 v_{\C}^{\step}(x,vd \gg 1) = \frac{I_L-I_R}{1+\exp{\{2[(\sqrt{2I_L}-\sqrt{2I_R})(x+x_0^{'})]}\}} \;,
 \label{step*}
\end{equation}
with
\begin{eqnarray}
 x_0^{'}(N) &=& \frac{1}{2(\sqrt{I_L}-\sqrt{I_R})} \nonumber \\
        &\times&  \left[ d(\sqrt{I_L}+\sqrt{I_R}) - \frac{1}{\sqrt{2}}
        \ln\left(\frac{\sqrt{I_R}(N-1)}{\sqrt{I_L}} \right) \right] \; . \nonumber \\
 \; .
 \label{x0'}
\end{eqnarray}
Using once more the parameters of Fig.~4, one obtains $x_0'(1.2) \sim 12.62$, $x_0'(2) \sim 12.43$. 
As before, $v_{\C}^{\barr}(x,vd \gg 1)$ presents a maximum at $x = - \, x_0'(N)$, while $v_{\C}^{\step}(x,vd \gg 1)$
vanishes exponentially for $x < - \, x_0'(N)$, and increases monotonically towards its limiting value $I_L-I_R$
for $x > -\, x_0'(N)$. Interestingly, the sum of both potentials also increases monotonically around $-\,x_0'(N)$.

\begin{figure}[h]
\begin{center}
\includegraphics[width=8.6cm]{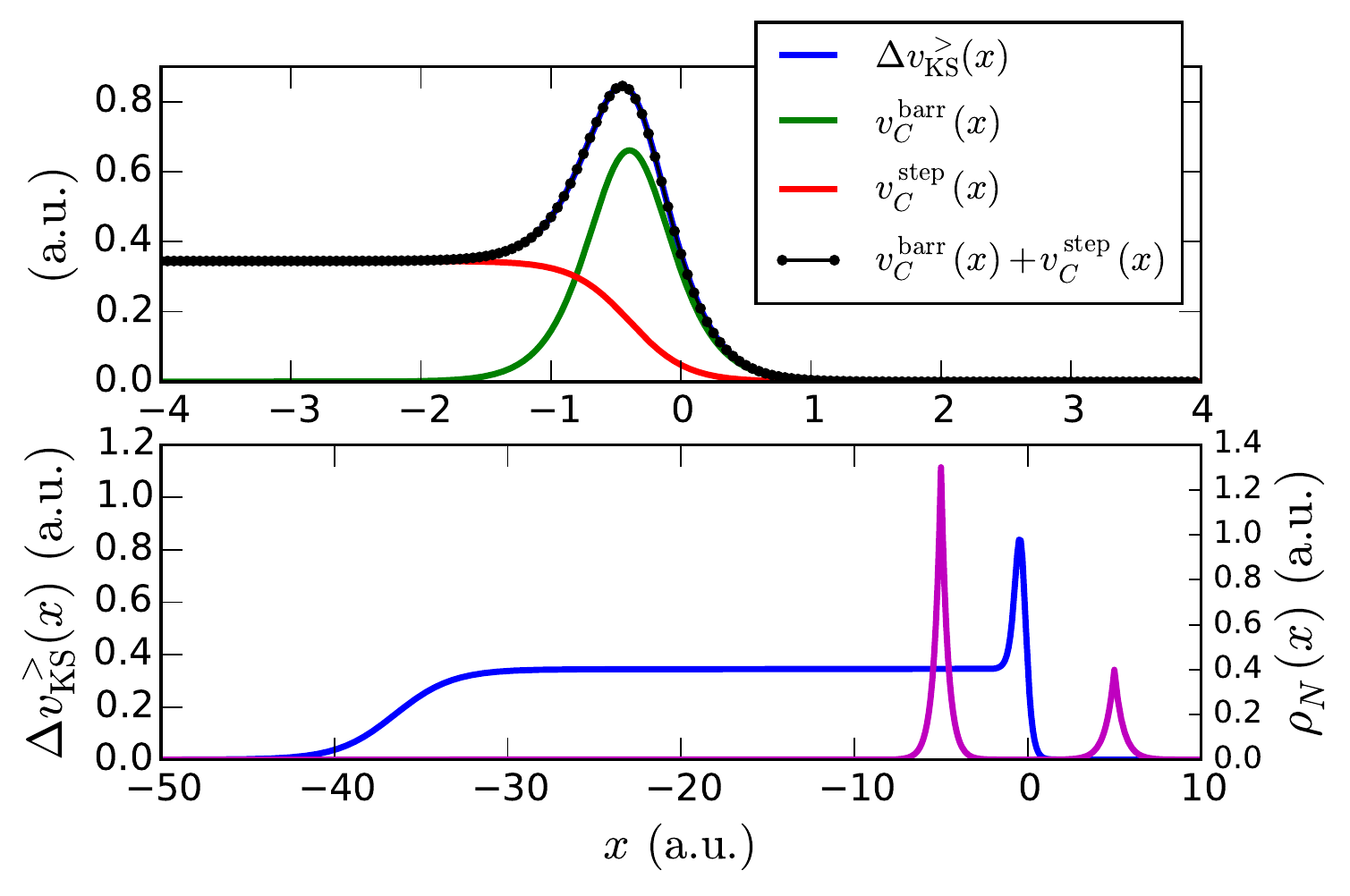}
\caption{$\Delta v_{\KS}^>(x)$ for $v=1$, $\lambda=1.3$, $N=1.4$, and $d=10$. Top panel: detail of the shoulder/barrier
contribution $v_{\C}^{\barr}(x)$ and the step contribution $v_{\C}^{\step}(x)$; lower panel: global view of 
$\Delta v_{\KS}^>(x)$,
and the associated density centered about $\pm d/2 = \pm 5$.
For these parameters, the plateau region has the height $I_L-I_R = 0.345$.}
\label{Fig5}
\end{center}
\end{figure}

We display in Fig.~5 how the sum of the barrier and step contributions to $\Delta v_{\KS}^>(x)$ combines to yield the 
shoulder/barrier feature centered at $x = - \, x_0(1.4) \simeq - \, 0.39$. We remark that this shoulder/barrier
structure in the exact KS potential at the dissociation limit appears exactly at the position of the 
maximum overlap of the left and (effective) right density distributions, centered in this case at $x=-5$ and
$x=5$, respectively. The lower panel gives a global view of $\Delta v_{\KS}^>(x)$, showing the marked asymmetry of
$v_{\C}^{\step}(x)$, that although associated with the electron located at the left-well at $x=-5$, extends much
further towards the left side than towards the right side, taking as reference the left-well coordinate. From
Eq.~(\ref{x0'}), one obtains that $x_0'(1.4) \simeq 40$, in good agreement with the beginning of the left side
of $\Delta v_{\KS}^>(x)$.
The length of the plateau or step in $v_{\C}^{\step}(x)$ may be estimated from the difference
$x_0'(N) - x_0(N)$: it increases linearly with $d$, while display a logarithmic dependence on $N$,
decreasing its length as $N$ increases beyond 1.

A two-electron one-dimensional model of a heterodiatomic molecule composed of two-open shell atoms has been
also considered~\cite{TMM09}, with the Coulomb interaction being replaced by a soft-Coulomb potential. 
Their numerical results, restricted to the case $N=2$, are similar to ours, as they also
obtained the exact KS potential with a shoulder/barrier and step features.

\subsection{Further discussions}

\subsubsection{The dissociation of the $H_2^+$ molecule}

For $\lambda=1$ (homonuclear molecule), we have used as definition of the dissociation limit the condition 
$e^{-2vd} \ll 1$. For $\lambda \neq 1$ (heteronuclear molecule), on the other side, 
the condition for dissociation modifies to $2 \sqrt{\lambda} e^{-vd} \ll |\lambda-1|$. 
This suggest that for $\lambda \rightarrow 1$, for increasing $d$ the molecule may transition from a 
quasi-homonuclear configuration to a quasi-heteronuclear configuration; the situation is particularly interesting
for the case of $N \rightarrow 1^+$. The critical value of $d$ for that
transition is approximately given by the equation $e^{-vd^*} = |\lambda-1|/(2\sqrt{\lambda})$. Solving for $d^*$, it yields
\begin{equation}
 d^* \simeq - \frac{1}{v} \ln\left( \frac{|\lambda-1|}{2\sqrt{\lambda}} \right) \; .
 \label{d*}
\end{equation}
If $\lambda \rightarrow 1$ from above or from below, $d^* \rightarrow \infty$ and the molecule is in the 
homonuclear limit for any finite value of $d$. Writing $\lambda=1+\epsilon$, with $\epsilon \ll 1$, 
$d^*(\lambda=1+\epsilon) \sim -\ln|\epsilon|/v$;
for $d \lesssim d^*$, the molecule (a one-dimensional mimic of the $H_2^+$ molecule if
$N \rightarrow 1^+$)
will display a density distribution essentially of the homonuclear case, which on the other side is expected since
$\lambda \sim 1$. But for $d \gtrsim d^*$, the density will start to display instead a quasi-heteronuclear 
configuration, with electrons being transferred from the atom with the lower ionization potential towards the 
one with the higher ionization potential. This is not as intuitive as in the previous case, since in
principle $\lambda \sim 1$ is associated with a ``bonding'' charge distribution.
This explains why the homonuclear limit of Eq.~(\ref{barr}) does not coincide with the strict homonuclear result
of Eq.~(\ref{vbarr}): for arriving to the former we have assumed that the system is in a heteronuclear configuration
such that $d \gtrsim d^*$, and after this is not possible to recover the homonuclear limit with $d \lesssim d^*$.
The results displayed in Figs.~4 and 5 are well inside the heteronuclear regime, considering that $d^* \simeq 1.86$
and $d^* \simeq 2.03$ in these two figures, respectively.

\begin{figure}[h]
\begin{center}
\includegraphics[width=8.6cm]{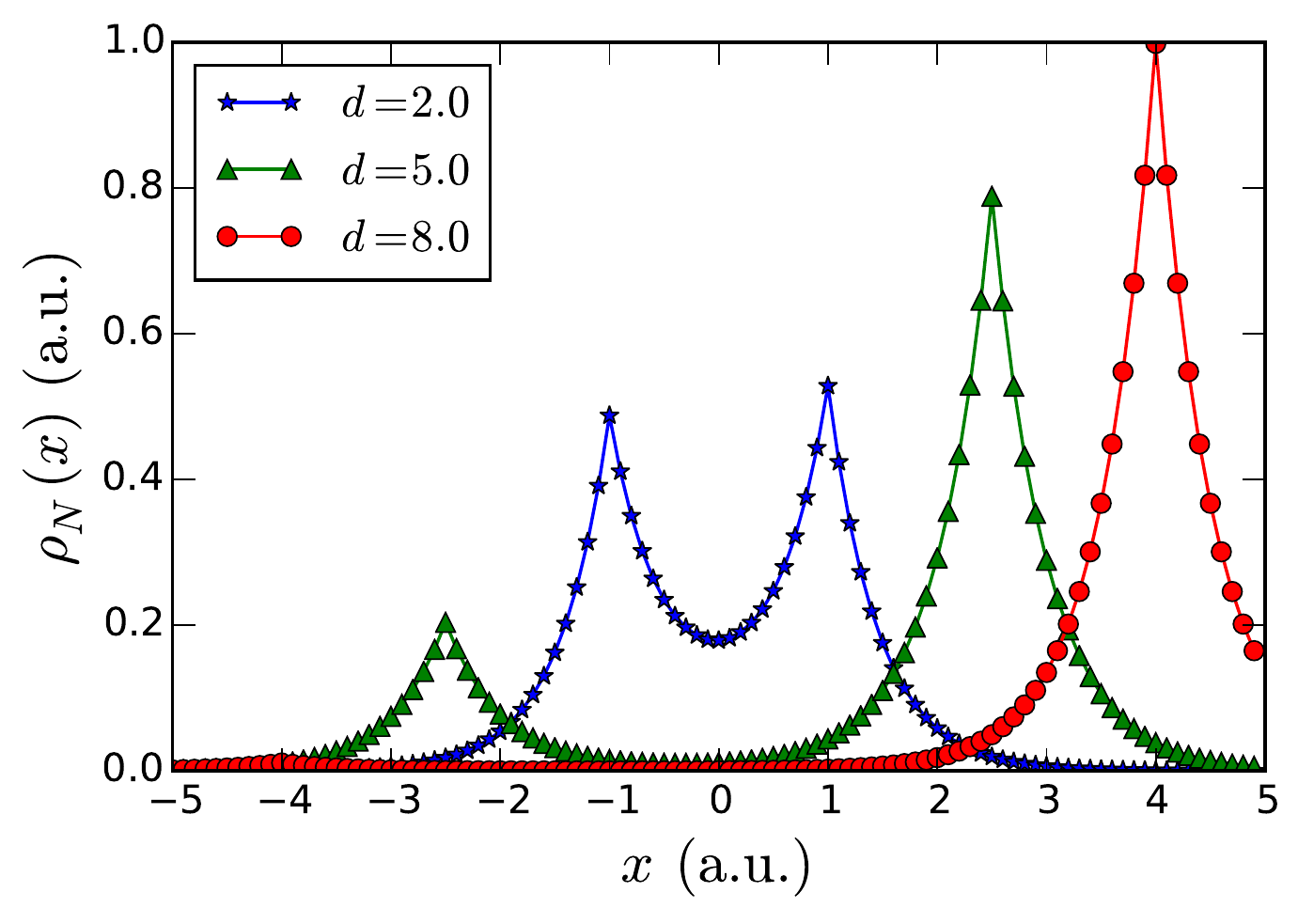}
\caption{Interacting density versus coordinate $x$, for different molecule's sizes. $v=1, \lambda = 0.99$, and $N=1.01$.
The configuration transition homonuclear $\rightarrow$ heteronuclear is evident as $d$ passes through
$d^* \sim 5.3$.}
\label{Fig6}
\end{center}
\end{figure}

We display in Fig.~6 an example of this situation. For these parameters, $d^* \sim 5.3$ from Eq.~(\ref{d*}).
Note that this dramatic change in the molecule's electronic density from a $H^{+0.5} \cdots \; H^{+0.5}$-like
configuration to a $H^{+1} \cdots H $-like configuration is achieved through the ``step'' contribution
to $\Delta v_{\KS}^>(x)$, whose height $I_R-I_L \simeq v(1-\lambda^2) \simeq 0.02$ is very small, and can be as small as
desired by choosing $\lambda$ closer to 1.

The limit $N \rightarrow 1^+$
of these results suggest a simple intuitive scenario for the ``physical'' dissociation process of the real
tridimensional $H_2^+$ molecule. As the separation between the two protons increases beyond the equilibrium
distance, the molecule will start to feel increasingly the effect of any of the symmetry breaking fields from the 
environment (represented in our model by having $\lambda \neq 1$), and no matter how small this breaking field may
be, charge will be transfered from one proton to the other about some critical distance $d^*$. 
After this, and by increasing $d$ further, the probability for the molecule to return to the symmetric charge distribution will become extremely
small, as the tunneling probability for this process decreases exponentially as $d$ increases.

\subsubsection{The anomalous scaling of $v_{\KS}(x)$ with the contact repulsion parameter $\gamma$}

It is also of some fundamental interest to analyze how the KS potential scales with $\gamma$. According to the ensemble generalization
of the KS formulation of ground-state DFT~\cite{PPLB82},

\begin{equation}
 v_{\KS}^N(x) = v_{\ext}(x) + v_{\HXC}([\rho_N];\gamma) \; ,
 \label{KSp}
\end{equation}
with
\begin{equation}
 v_{\HXC}([\rho_N];\gamma) = v_{\H}([\rho_N];\gamma) + v_{\X}([\rho_N];\gamma) + v_{\C}([\rho_N];\gamma) \; .
\end{equation}
$v_{\H}([\rho_N];\gamma)$, $v_{\X}([\rho_N];\gamma)$, $v_{\C}([\rho_N];\gamma)$ are the Hartree, exchange, and correlation
contributions to the KS potential, respectively. They depend on the contact parameter $\gamma$. For example,
within our model
\begin{equation}
 v_{\H}([\rho_N];\gamma) = \int dx' \rho_N(x') \, \gamma \, \delta(x-x') = \gamma \, \rho_N(x) \; .
\end{equation}
Besides, $v_{\X}([\rho_N];\gamma) = - \, v_{\H}([\rho_N];\gamma)$ for $0 \leq N \leq 1$, while~\cite{GGS09}
\begin{equation}
 v_{\X}([\rho_N];\gamma) = - \frac{N^2-2N+2}{N^2} v_{\H}([\rho_N];\gamma) + \frac{2(2-N)}{N^3}U[\rho_N] \; ,
\end{equation}
for $1 < N \leq 2$. Here,
\begin{eqnarray}
 U[\rho_N] &=& \frac{1}{2} \int dx \int dx' \rho_N^>(x) \rho_N^>(x') \, \gamma \, \delta(x-x') \nonumber \\
           &=& \frac{\gamma}{2} \int dx [\rho_N^>(x)]^2 \; .
\end{eqnarray}
The problem here is that for $\gamma \rightarrow \infty$ (our case), 
$v_{\H}([\rho_N];\gamma)$, $v_{\X}([\rho_N];\gamma)$, and $v_{\C}([\rho_N];\gamma)$ all diverge linearly with $\gamma$.
On the other side, our previous results are such that both $\Delta v_{\KS}^<(x)$ and $\Delta v_{\KS}^>(x)$
are {\it finite}. How is this possible?.

The answer is clear for $0 \leq N \leq 1$: here $v_{\H}([\rho_N];\gamma) + v_{\X}([\rho_N];\gamma) \equiv 0$, canceling mutually,
whatever the value of $\gamma$. $v_{\C}([\rho_N];\gamma) \equiv 0$ also, since for $N \leq 1$ the system is not correlated.
This is of course consistent with the result $\Delta v_{\KS}^<(x) \equiv 0$ obtained before.
The situation for $1 < N \leq 2$ is more interesting, since in this case the cancellation of the divergent
contributions in $v_{\H}([\rho_N];\gamma)$ and $v_{\X}([\rho_N];\gamma)$ is only partial. 
For $v_{\HXC}([\rho_N];\gamma)$ in Eq.~(\ref{KSp})
to remain finite, one concludes that a necessary condition is that $v_{\C}([\rho_N];\gamma)$ must be the sum of two 
contributions: $v_{\C}([\rho_N];\gamma) = \tilde{v}_{\C}([\rho_N];\gamma) + \Delta v_{\KS}^{>}(x)$. 
$\tilde{v}_{\C}([\rho_N];\gamma)$ 
must scale linearly with $\gamma$, and should fulfill the cancellation constraint 
$v_{\H}([\rho_N];\gamma) + v_{\X}([\rho_N];\gamma) + \tilde{v}_{\C}([\rho_N];\gamma) \equiv 0$, while the 
second contribution,
which remains finite even when $\gamma \rightarrow \infty$ is what we have denoted as
$\Delta v_{\KS}^>(x)$ in Eq.~(\ref{Cp}). In the more general case of finite $\gamma$ 
or by using more realistic 
(Coulomb-like) potentials, we expect that both 
$v_{\H}([\rho_N];\gamma) + v_{\X}([\rho_N];\gamma) + \tilde{v}_{\C}([\rho_N];\gamma)$ and
$\Delta v_{\KS}^>(x)$ will be different from zero~\cite{note}.
Our strongly interacting model exhibits somehow this anomalous scaling in its extreme limit,
forcing the complete mutual cancellation of all the ``normal'' scaling contributions to $v_{\HXC}([\rho_N];\gamma)$,
and keeping finite only the ``anomalous'' contribution $\Delta v_{\KS}^>(x)$, that is of zero-order in $\gamma$.

We expect then that the main result of this section regarding the anomalous scaling property of the KS potential
to remain valid even after the replacement of the delta-function interaction potential by more realistic
models~\cite{note,YBLVGG15}.

The fact that the Kohn-Sham potential must include a term with an anomalous scaling in the strength of the 
Coulomb interaction has been recently noticed~\cite{YBLVGG15}, and our present results confirm even more clearly this important point.
The conclusion is clear: without these anomalous scaling terms, the dissociation limit of homonuclear and heteronuclear
molecules will be poorly described, since the KS potential will suffer from the absence of the barrier and step
features in it. As discussed in Ref.~[\onlinecite{YBLVGG15}], the absence of these ``Mott barriers'' has also important
consequences on the electronic properties of strongly correlated solids.

\section{Conclusions}
Some unusual features of the exact Kohn-Sham potential for finite systems with a fractional number of electrons
have been discussed. The exact ground-state density of a strongly interacting model for a one-dimensional diatomic
molecule has been obtained, from which by reverse engineering the Kohn-Sham potential is derived for all fractional
molecules occupancies between 0 and 2. Large differences exist between the results for the homonuclear and
heteronuclear cases, particularly in the dissociation limit.
For the homonuclear case, and in the dissociation limit, the exact Kohn-Sham potential
builds a barrier at the midpoint between the two atoms, whose strength increases linearly with $N$, with
$1 < N \leq 2$. In the heteronuclear case, the KS potential exhibits a peak related to the barrier of the 
homonuclear case, exactly centered at the coordinate of maximum overlap between the left and right density 
distributions. Besides, it also displays a plateau around the atom with
the higher ionization potential, whose size (but not its strength) depends on $N$.
An anomalous zero-order scaling of the KS potential with regards to the strength of the electron-electron
repulsion is clearly displayed by our model, without which both the unusual barrier and plateau features
will be absent.

\begin{acknowledgements}
The authors acknowledge J. Fuhr and G. Bocan for useful discussions, and to J. Luzuriaga for a critical reading of the manuscript. 
This work was sponsored by PICT 2012-0379 of the ANPCyT,
Argentina. C.R.P. is partially supported by CONICET.
\end{acknowledgements}


\begin{thebibliography}{9}
\bibitem{YG05} V. I. Yukalov and M. D. Girardeau, Laser Phys. Lett. \textbf{2}, 375 (2005). In a few words,
the ground-state of the diatomic molecule for $N=2$ is a spin-singlet. Since the total wave-function must
be antisymmetric under the exchange of the two electrons, this implies that the spacial part of it should
be symmetric (like bosons). On the other side, for $\gamma \rightarrow \infty$ the contact potential imposes
the vanishing condition $\Psi^{\gamma \rightarrow \infty}(x,x)=0$. 
Eq.~(\ref{bfm}) is then the exact expression for the spatial
part of the ground-state $N=2$ electronic wavefunction that is both symmetric, and vanishes when the two 
electronic coordinates coincide.
\bibitem{convexity} For Eq.~(\ref{density*}) to be valid, the total energy must satisfy the ``convexity''
 condition $E(N+1)+E(N-1)>2E(N)$. Taking $N=1$, this implies the condition 
 $E(2)=\varepsilon_+ + \varepsilon_- > 2 E(1) = 2 \varepsilon_+$, which is strictly satisfied, since
 $\varepsilon_- > \varepsilon_+$.
\bibitem{note*} This is easily seen by using the starting equation for the reverse engineering procedure,
Eq.~(\ref{KSp*}): for $\rho_N^>(x) \sim N[\rho_L(x)+\rho_R(x)]$, the factor $N$ cancels.
\bibitem{note} We have partially confirmed this by replacing the contact potential $\gamma \, \delta(x-x')$ by a
``soft'' one-dimensional Coulomb potential $1/(|x-x'|+a)$, with $a$ being a small positive constant.
In the limit $a \ll 1$, we have checked: a) the nearly complete cancellation of the ``normal'' contributions
to $v_{Hxc}(x)$ for any $x$; and b) the existence of an anomalous contribution to $v_{Hxc}(x)$, whose size does not
depends on $a$, and that remains finite whatever the infinitesimal value of $a$ is.
\end{thebibliography}
\end{document}